\documentclass[11pt]{article}
\usepackage{iap2000,epsfig,rotate}

\def\be{\begin{equation}}
\def\ee{\end{equation}}
\def\bea{\begin{eqnarray}}
\def\eea{\end{eqnarray}}

\def\hmpc{~h$^{-1}$ Mpc~}

\begin{document}
\vspace*{4cm}
\title{ LARGE-SCALE DISTRIBUTION and SPECTRAL PROPERTIES
        of GALAXIES in the SHAPLEY CONCENTRATION }

\author{ ELENA ZUCCA \& SANDRO BARDELLI }
\address{ Osservatorio Astronomico di Bologna, via Ranzani 1, 
          I-40127 Bologna, Italy \\
          {\rm A.BALDI, G.ZAMORANI, L.MOSCARDINI, R.SCARAMELLA } }

\maketitle\abstracts{
We present the results of a redshift survey of both cluster and intercluster
galaxies in the central part of the Shapley Concentration, the richest
supercluster of clusters in the nearby Universe, consisting of $\sim 2000$
radial velocities.
We estimate the total overdensity in galaxies of the supercluster and its mass
and we discuss the cosmological implications of these results. 
Moreover, using a Principal Components Analysis technique, we study the 
influence of the cluster and supercluster dynamics on the galaxy spectral 
morphology. 
}

%
\noindent
Clusters of galaxies are thought to form by accretion of subunits in a 
hierarchical bottom up scenario: numerical simulations revealed that mergings 
happen along preferential directions, called density caustics, which define 
matter flows, at whose intersection rich clusters are formed.
\\
The central regions of rich superclusters, where the overdensity is of the 
order of $\sim 10$, are thought to be in the early collapse phase and 
therefore give the possibility to study the formation of clusters and groups 
of galaxies. 
These regions are the ideal environment for the detection of 
cluster mergings, because the peculiar velocities induced by the enhanced 
local density of the large scale structure favour the cluster-cluster and 
cluster-group collisions, in the same way as the caustics seen in the 
simulations. 
\\
Spectacular major merging events are visible in the central region of the
Shapley Concentration, the richest supercluster of galaxy clusters in the
nearby Universe (Zucca et al. 1993). Particularly interesting are two
chains of interacting clusters: the A3558 and the A3528 complexes (Fig. 1). 
\\
We are carrying on a long term multiwavelength study on the Shapley 
Concentration (see also the poster by Bardelli et al., these proceedings), 
in order to determine its mass, dynamical state and cosmological relevance, 
as well as to analyse in detail the physics of the merging phenomena. 
\\
In this poster we focus on the large-scale distribution and spectral
properties of galaxies both in clusters and in the ``field" in between
clusters. 

\begin{figure}
\begin{center}
\epsfxsize=0.7\hsize
\rotate[r] { \epsfbox {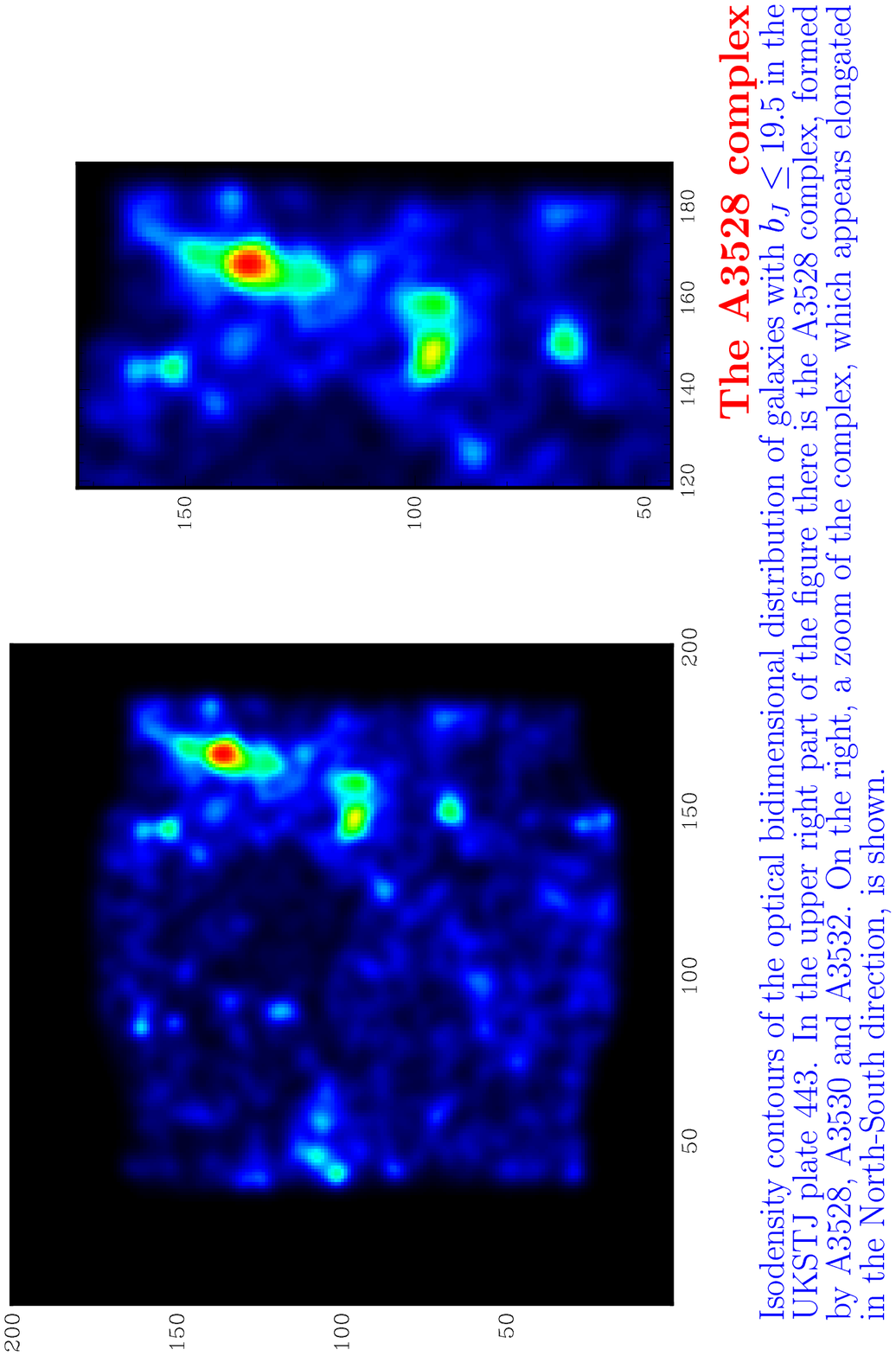} }
\epsfxsize=0.7\hsize
\rotate[r] { \epsfbox {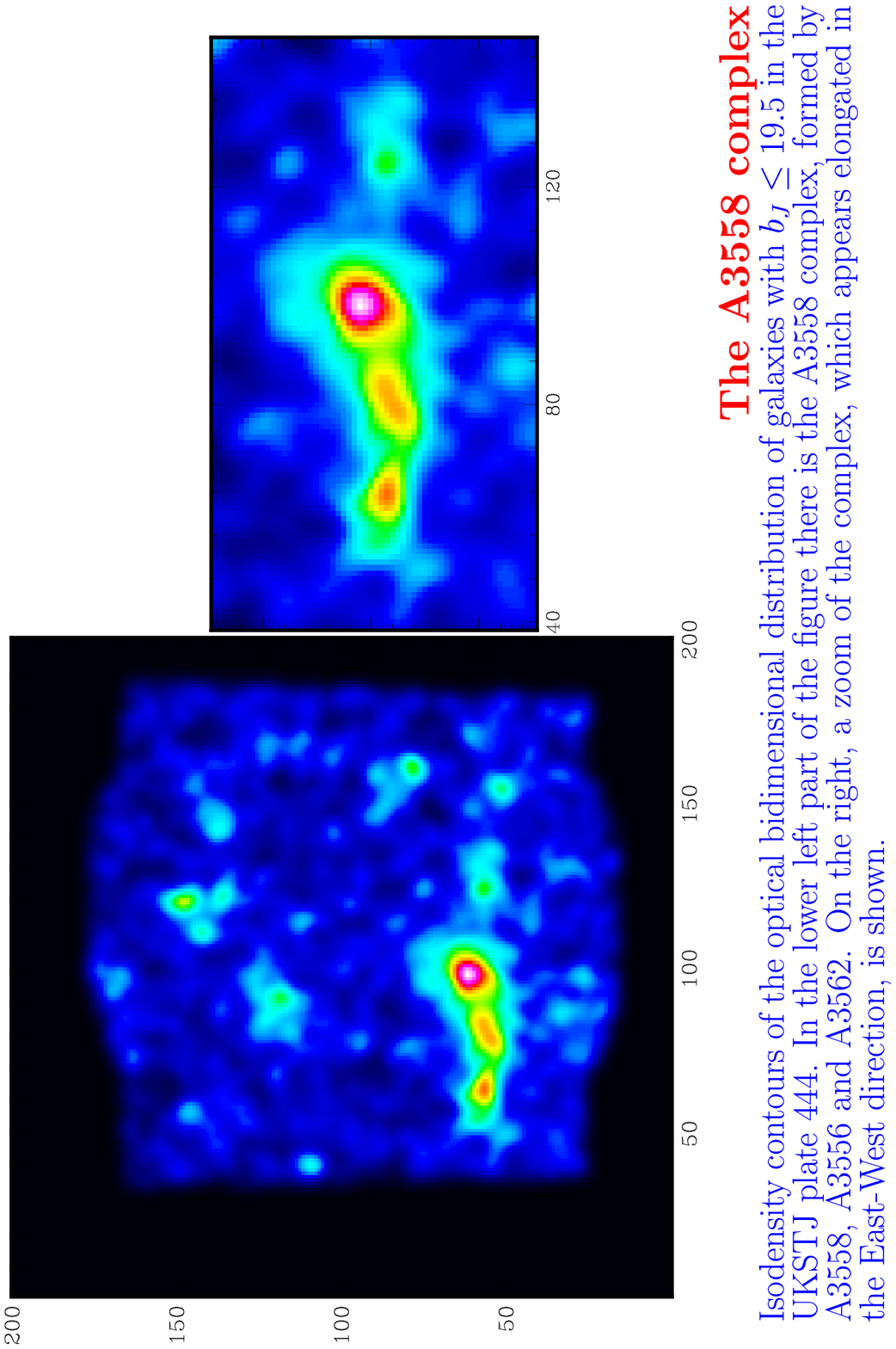} }
\end{center}
Figure 1: Optical distribution of galaxies in the central part of the
Shapley Concentration. 
\end{figure}

%
\section{Large-scale distribution of galaxies}
{\bf In collaboration with G.Zamorani, L.Moscardini, R.Scaramella}
\\
The distribution of clusters inside the Shapley Concentration is quite well 
studied; on the contrary, little is known about the distribution and the 
overdensity of intercluster galaxies. 
\\
For this reason we performed a redshift survey of galaxies which are outside
the clusters and the obvious bi--dimensional overdensities in this region,
in order to derive the distribution and the overdensity of ``field" objects. 
These data have been added to our surveys covering the cluster complexes
(Bardelli et al. 1994, 1998a, 2000b), resulting in a total redshift sample of 
$\sim 2000$ radial velocities.
In Figure 2 we show our sampling strategy: big and small circles correspond to
the field of view of the MEFOS and OPTOPUS multifiber spectrographs (at the 
3.6m ESO telescope), respectively. 
\\
In Figure 3 we show the wedge diagram of our sample. It is clear the presence
of the two cluster complexes dominated by A3558 (on the left) and
by A3528 (on the right). These two structures appear to be connected by a 
bridge of galaxies, resembling the Coma-A1367 system, the central
part of the Great Wall. The scale of this system is $\sim 23$ \hmpc and it is
comparable to that of Coma-A1367 ($\sim 21$ \hmpc). Note the presence
of two voids at $\sim 20000$ km/s and $\sim 30000$ km/s in the 
easternmost half of the wedge, labelled as V1 and V2 respectively. 
Other two voids (V3 and V4) are visible in the westernmost part of the plot:
in particular, void V3 appears to be delimited by two elongated features 
(S300a and S300b) which appear to ``converge" in a single feature at right 
ascension $\sim 13^h$ (S300c). We refer to this structure with the name S300: 
as shown below, the overdensity corresponding to this excess is remarkably 
high.
\\
Our main results are the following (Bardelli et al. 2000a):
\\
-- The average velocity of the observed intercluster galaxies in the Shapley
Concentration appears to be a function of the ($\alpha$, $\delta$) position,  
and can be fitted by a plane in the three--dimensional space ($\alpha$, 
$\delta$, $v$): the distribution of the galaxy distances around the best fit 
plane is described by a Gaussian with $\sigma =3.8$ \hmpc. 
\\ 
-- Using the 1440 galaxies of our sample in the magnitude range 
$17 - 18.8$, we reconstructed the density profile 
in the central part of the Shapley Concentration and we detected another 
significant overdensity at $\sim 30000$ km/s (see Figure 4).
\\
-- We estimate the total overdensity in galaxies, the mass and the dynamical
state of these structures, discussing the effect of considering a bias 
between the galaxy distribution and the underlying matter. 
The estimated total overdensity in galaxies of these two structures 
is $\displaystyle{ N \over \bar{N}}\sim 5.2$ on scale of $15.5$ \hmpc
for the Shapley Concentration and  $\displaystyle{ N \over \bar{N}} \sim 2.9$
on scale of $24.8$ \hmpc for S300. If light traces the mass distribution, 
the corresponding masses are 
$2.3\times 10^{16}$ $\Omega$ h$^{-1}$ M$_{\odot}$ and 
$5.1\times 10^{16}$ $\Omega$ h$^{-1}$ M$_{\odot}$ for Shapley Concentration
and S300, respectively (see Table 1 and Figure 5). 
\\
The dynamical analysis reveals that, if light traces 
mass and $\Omega=1$, the Shapley Concentration already reached its turnaround 
radius and started to collapse: the final collapse will happen in $\sim 3
\ 10^9$ h$^{-1}$ yrs. 
\\
-- We compare our mass estimates on various scales with other results in 
the literature, finding a general agreement (see Figure 5). 
\\
-- We find an indication that the value of the bias between clusters
and galaxies in the Shapley Concentration is higher than 
that reported in literature, confirming the impression that this supercluster
is very rich in clusters. 
\\
-- Finally from the comparison with some theoretical scenarios, we find that
the Shapley Concentration is consistent with the predictions of the models
with more power on large scale (such as Open and $\Lambda$ CDM), while it is 
inconsistent with the standard CDM normalized to the cluster abundance.

\begin{figure}
\begin{center}
\leavevmode
\epsfxsize=0.49\hsize \epsfbox{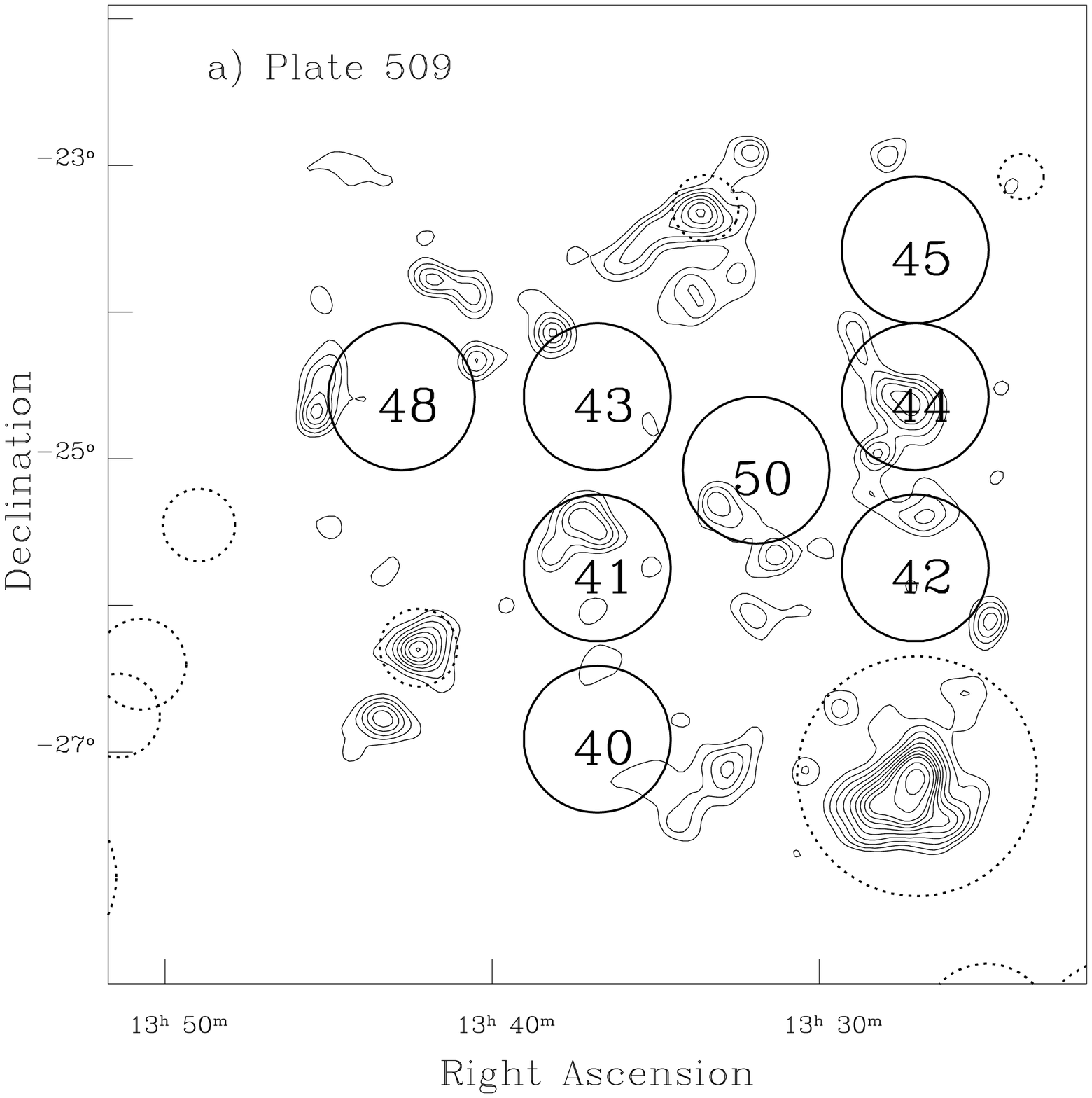} \hfil
\epsfxsize=0.49\hsize \epsfbox{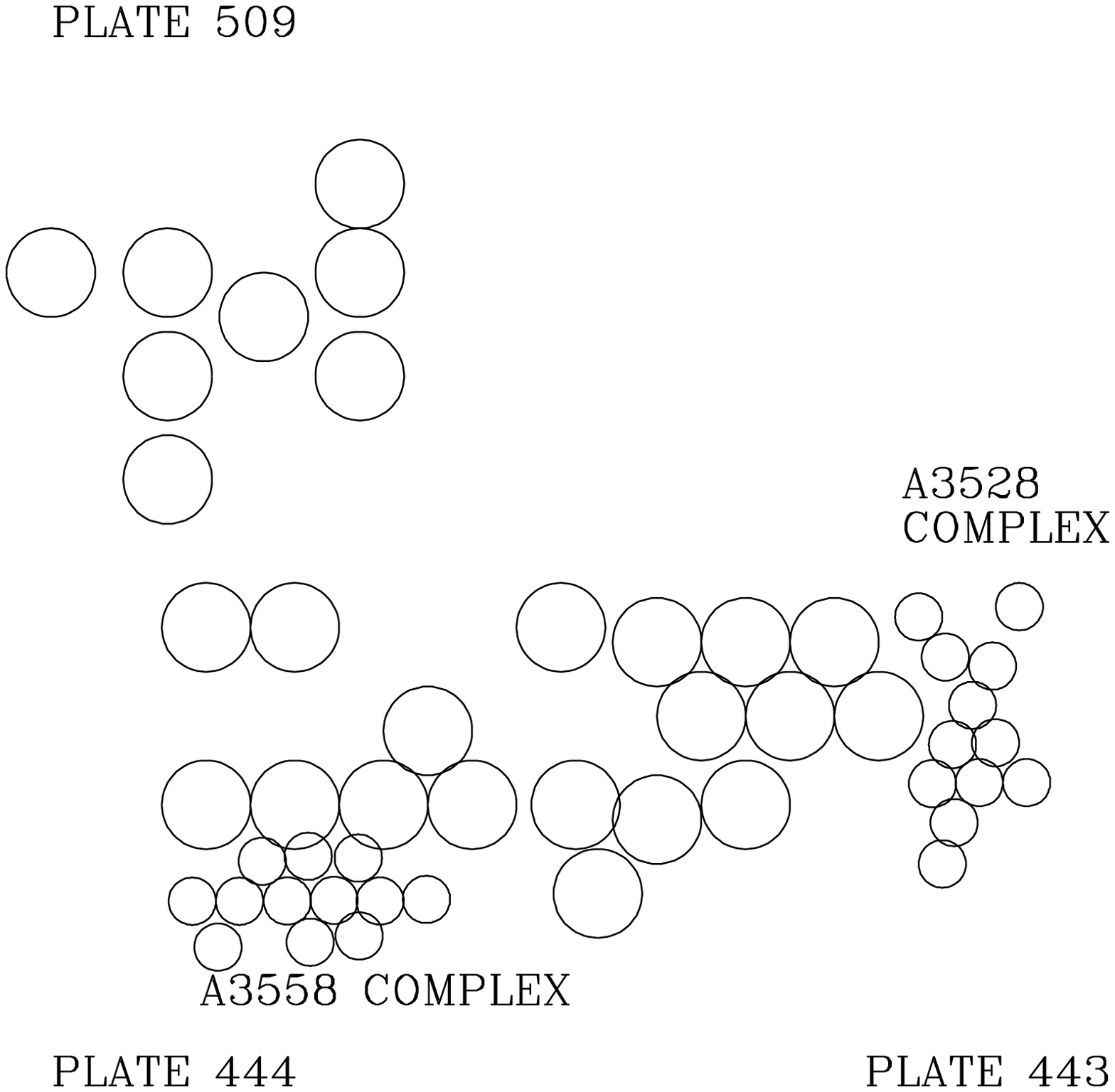} \hfil
\leavevmode
\epsfxsize=0.49\hsize \epsfbox{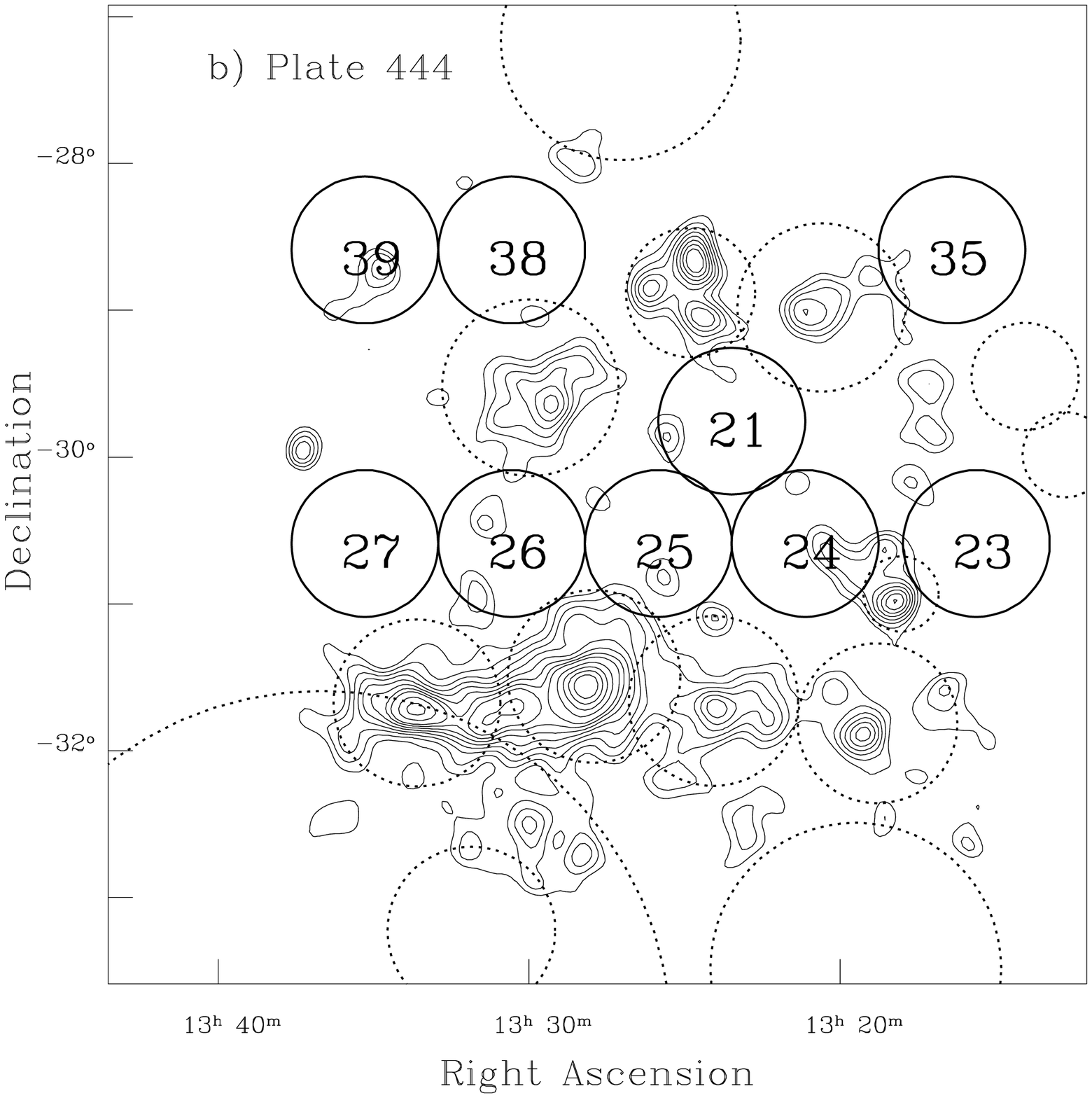} \hfil
\epsfxsize=0.49\hsize \epsfbox{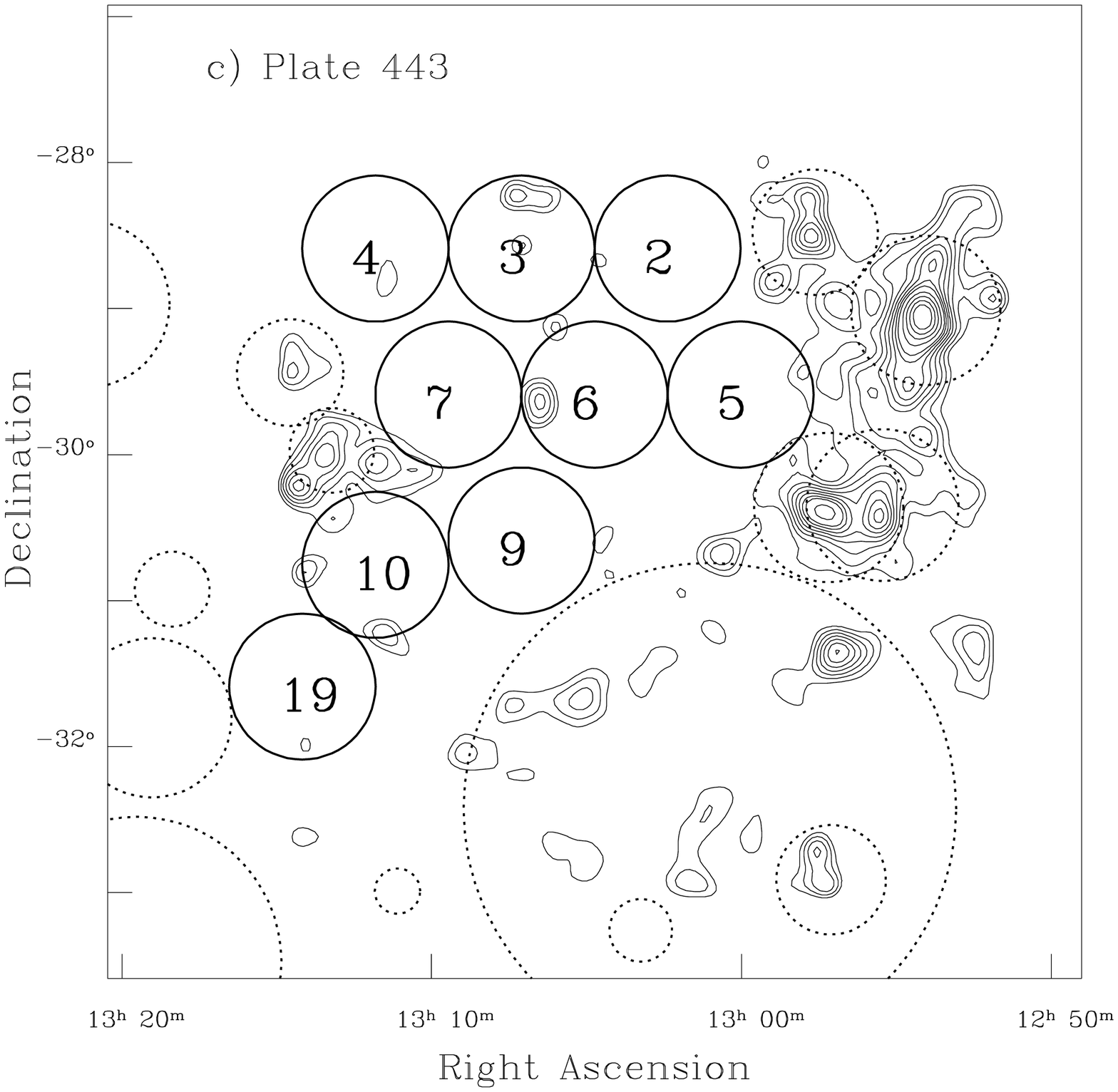} \hfil
\end{center}
Figure 2: Isodensity contours of the bidimensional distribution of the 
galaxies in the $b_J$ magnitude bin $17-19.5$ in the UKSTJ plates which
cover the central part of the Shapley Concentration. For the
Abell clusters present in the plates, circles of one Abell radius have been
drawn (dashed curves). Solid circles represent the MEFOS fields. 
\\
a) Plate 509; b) Plate 444; c) Plate 443. The two most evident systems are 
the A3558 (plate 444) and the A3528 (plate 443) cluster complexes. The dashed 
circle at the bottom right corner of plate 509 corresponds to the cluster 
A1736. 
\\ 
The relative positions of the fields is shown in the upper right panel, where
small circles represent OPTOPUS observations on the cluster complexes. 
\end{figure}

\begin{figure}
\epsfysize=11cm
\epsfbox{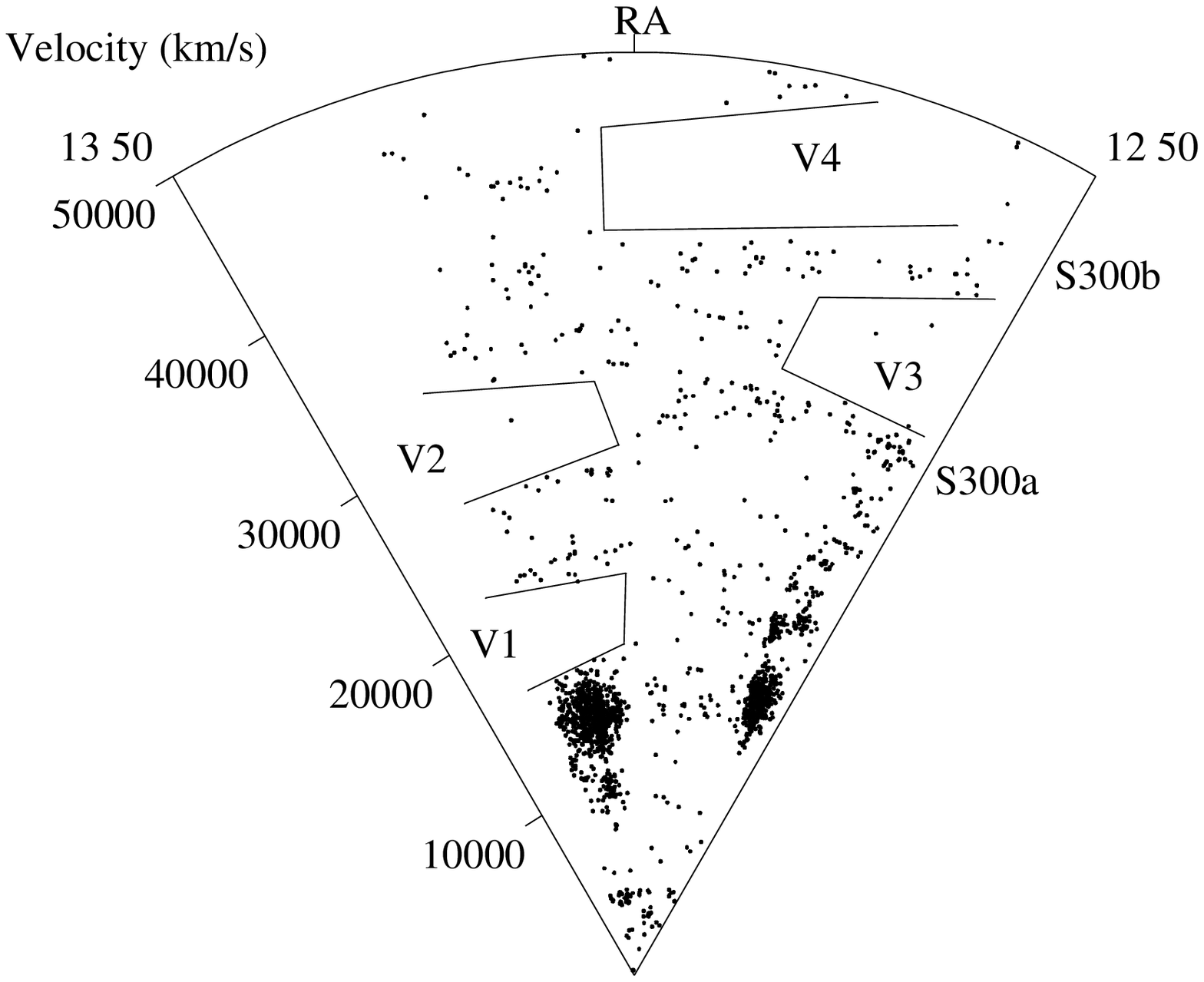}
\epsfysize=11cm
\epsfbox{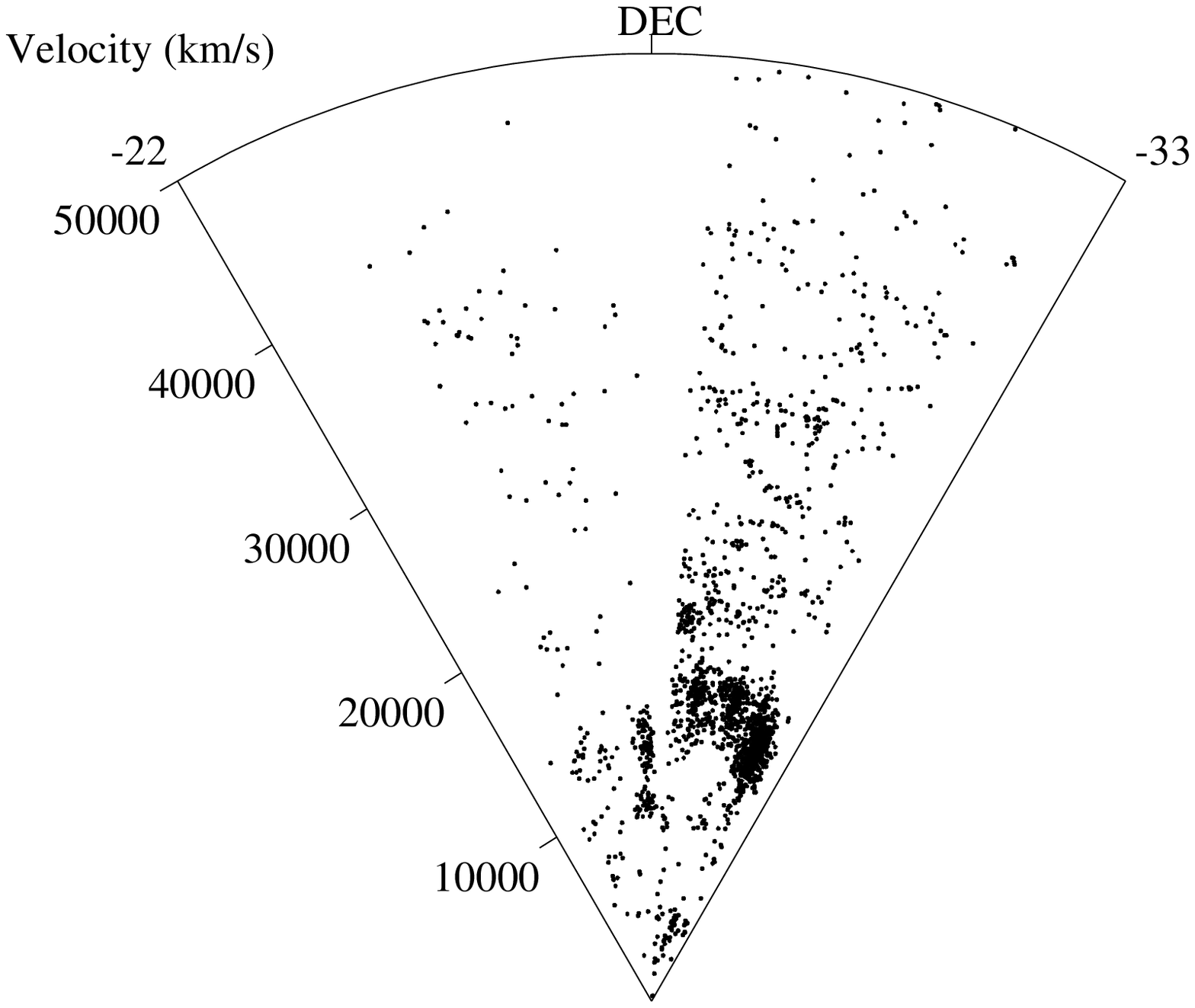}
Figure 3: Wedge diagram of the galaxy distribution in our samples.
Upper panel: 1728 galaxies in the plates 444 and 443 and in the velocity 
range $[0-50000]$ km/s. Note the clear presence of the A3558 and A3528 
complexes and of four voids. 
Lower panel: 1979 galaxies in plates 443, 444 and 509 in the velocity range 
[0-50000] km/s. Note that the plotted coordinate is the declination.
\end{figure}

\begin{figure}
\begin{center}
\leavevmode
\epsfxsize=0.49\hsize \epsfbox{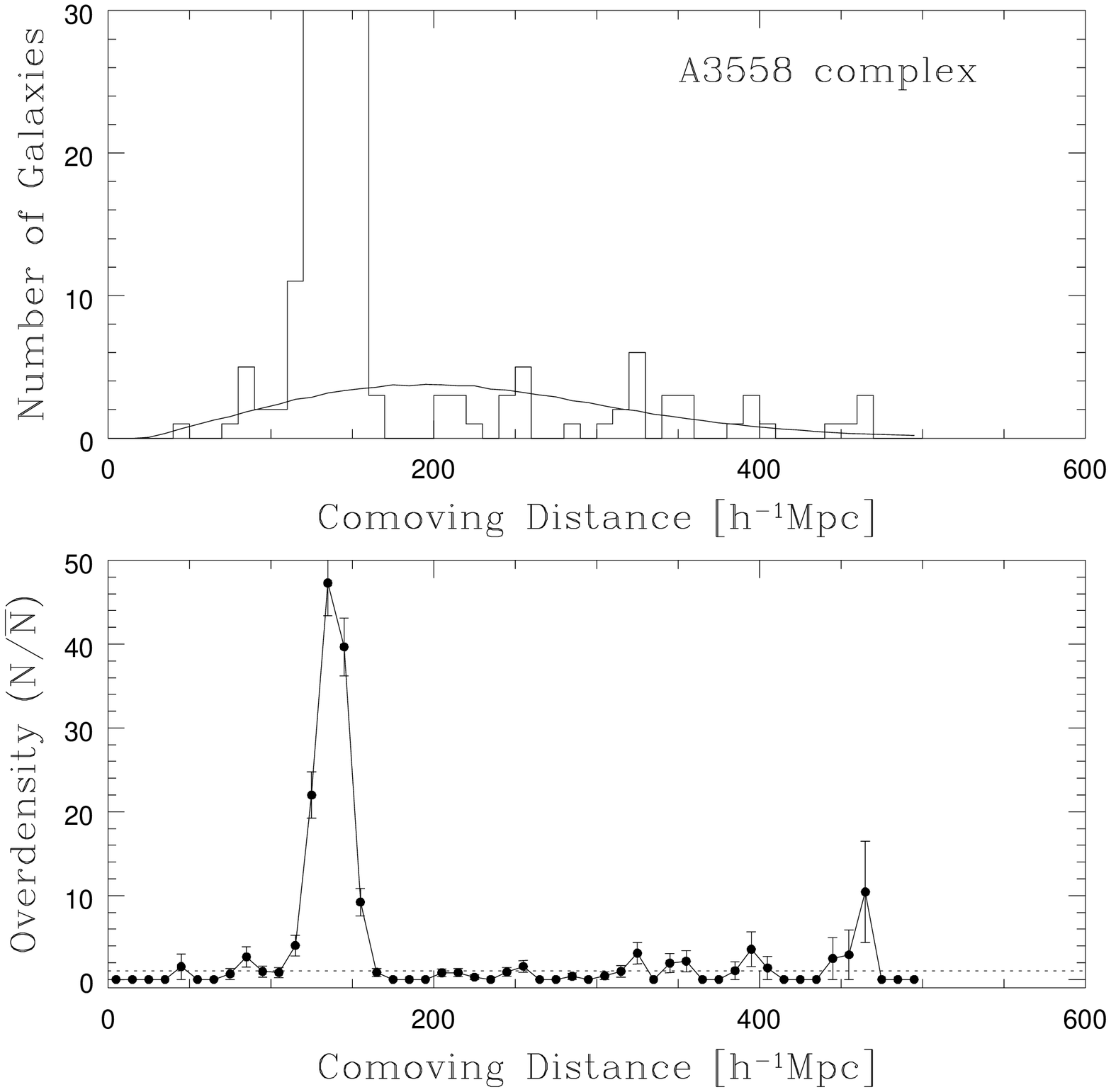} \hfil
\epsfxsize=0.49\hsize \epsfbox{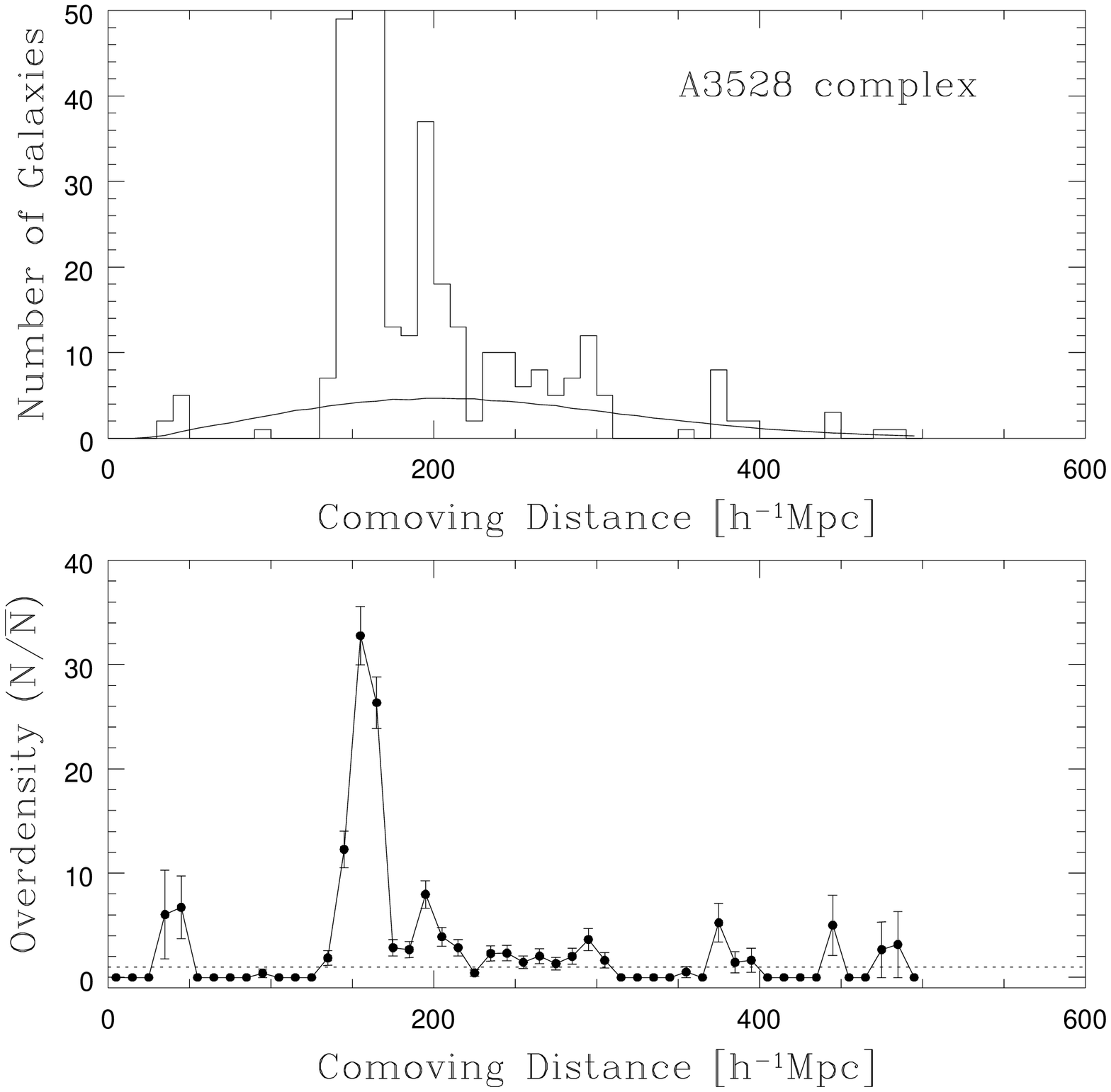} \hfil
\leavevmode
\epsfxsize=0.49\hsize \epsfbox{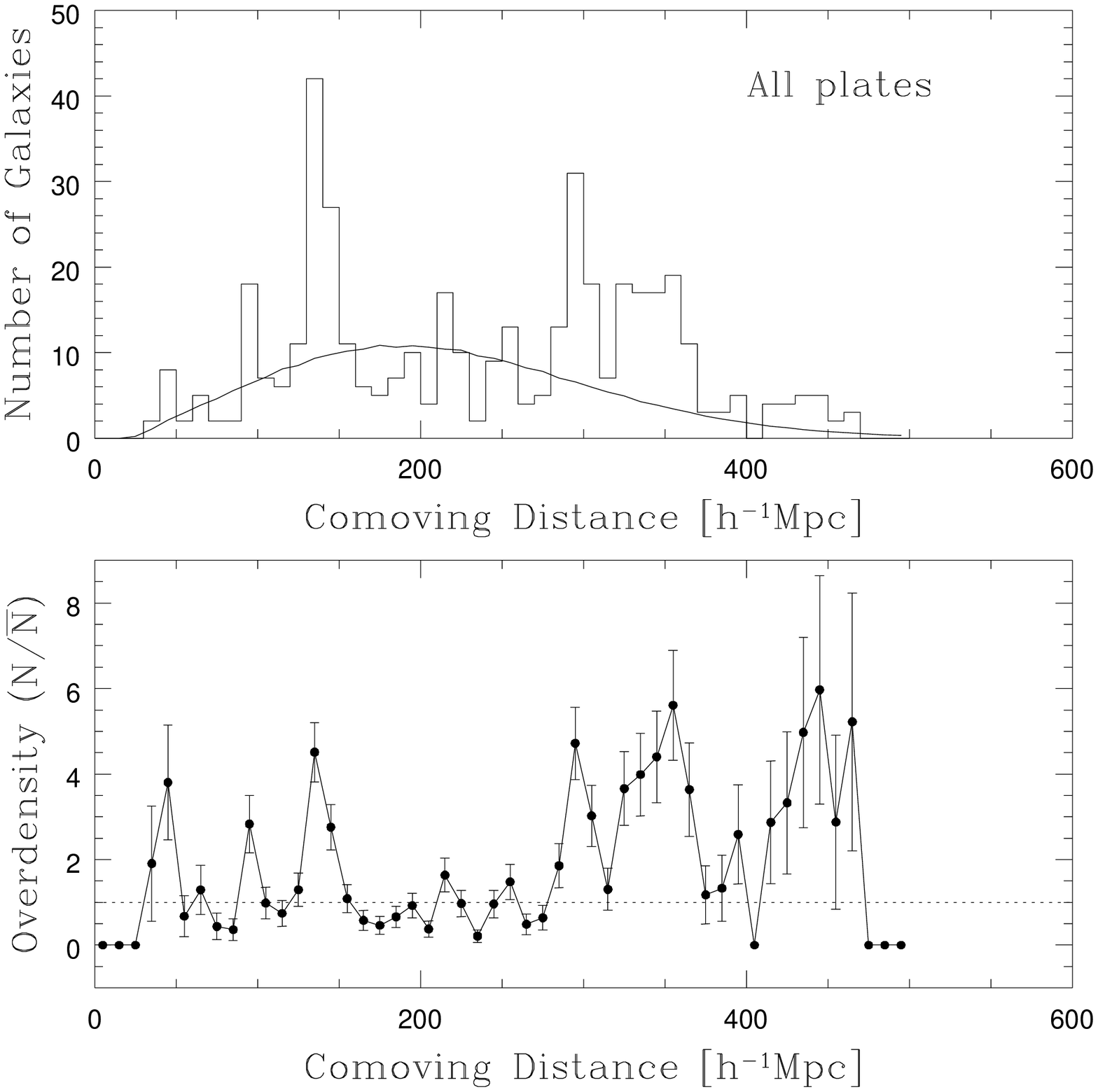} \hfil
\epsfxsize=0.49\hsize \epsfbox{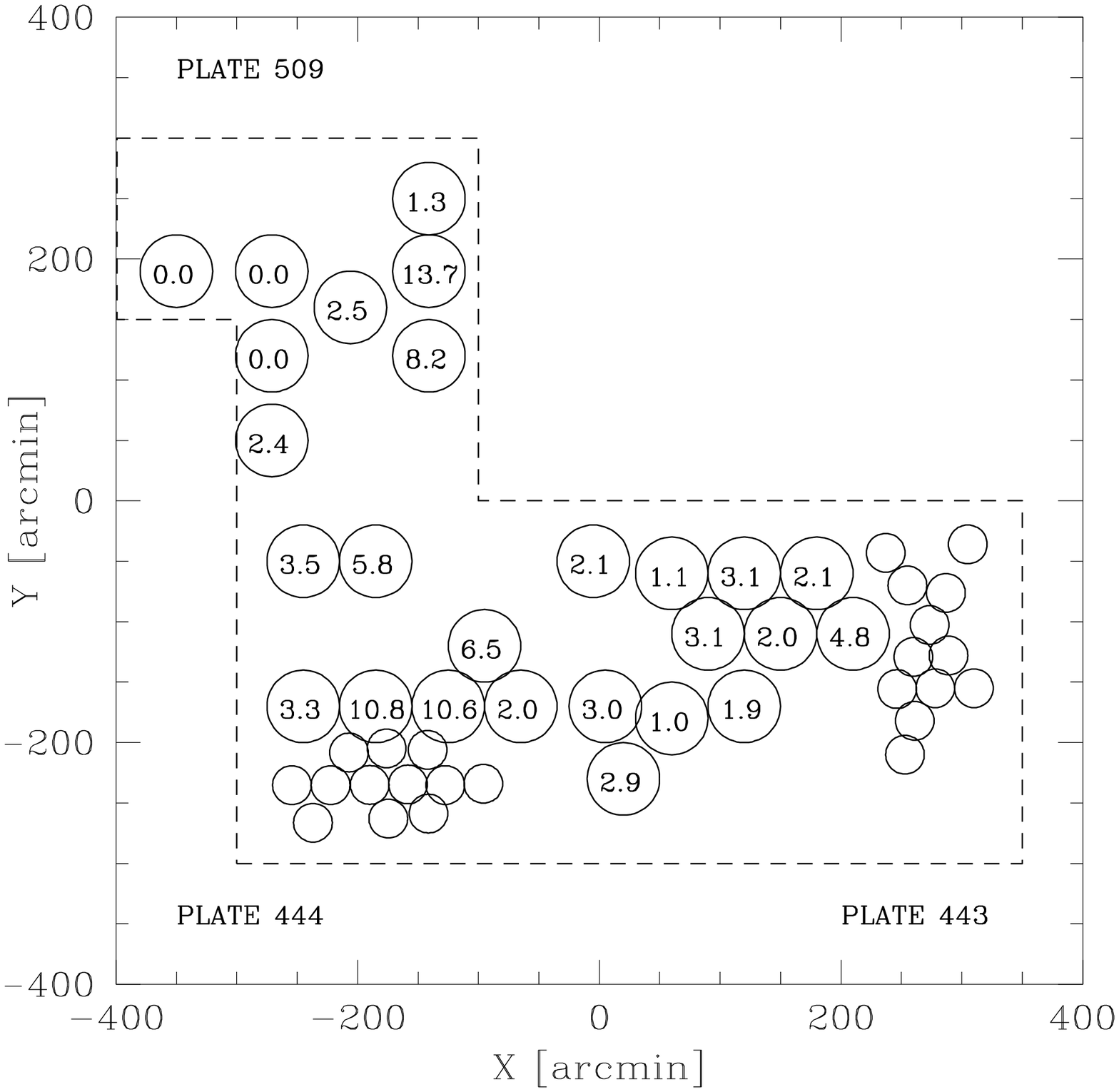} \hfil
\end{center}
Figure 4: 
Upper panels: Histograms of the velocities in the A3558 and A3528 complexes and 
in the intercluster sample, with superimposed the distribution expected for a
uniform sample. Lower panels: Galaxy density profiles. The dotted line
corresponds to $\displaystyle{ { N \over \bar{N}}  = 1}$.
The estimated overdensities in the single MEFOS fields as distributed in the 
sky are reported in the last figure. 
\end{figure}

\begin{figure}
\epsfysize=7cm
\begin{flushleft}
\epsfbox{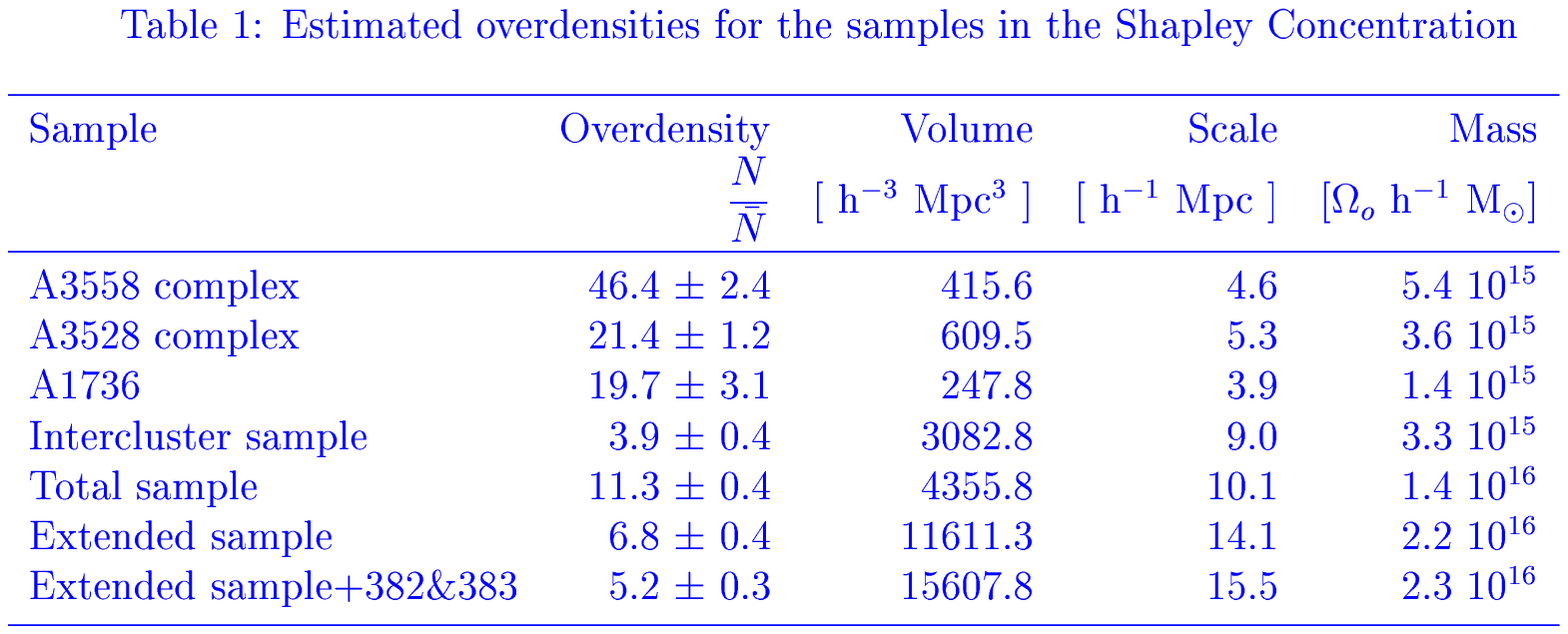} \hfil
\end{flushleft}
\epsfysize=11cm
\begin{center}
\epsfbox{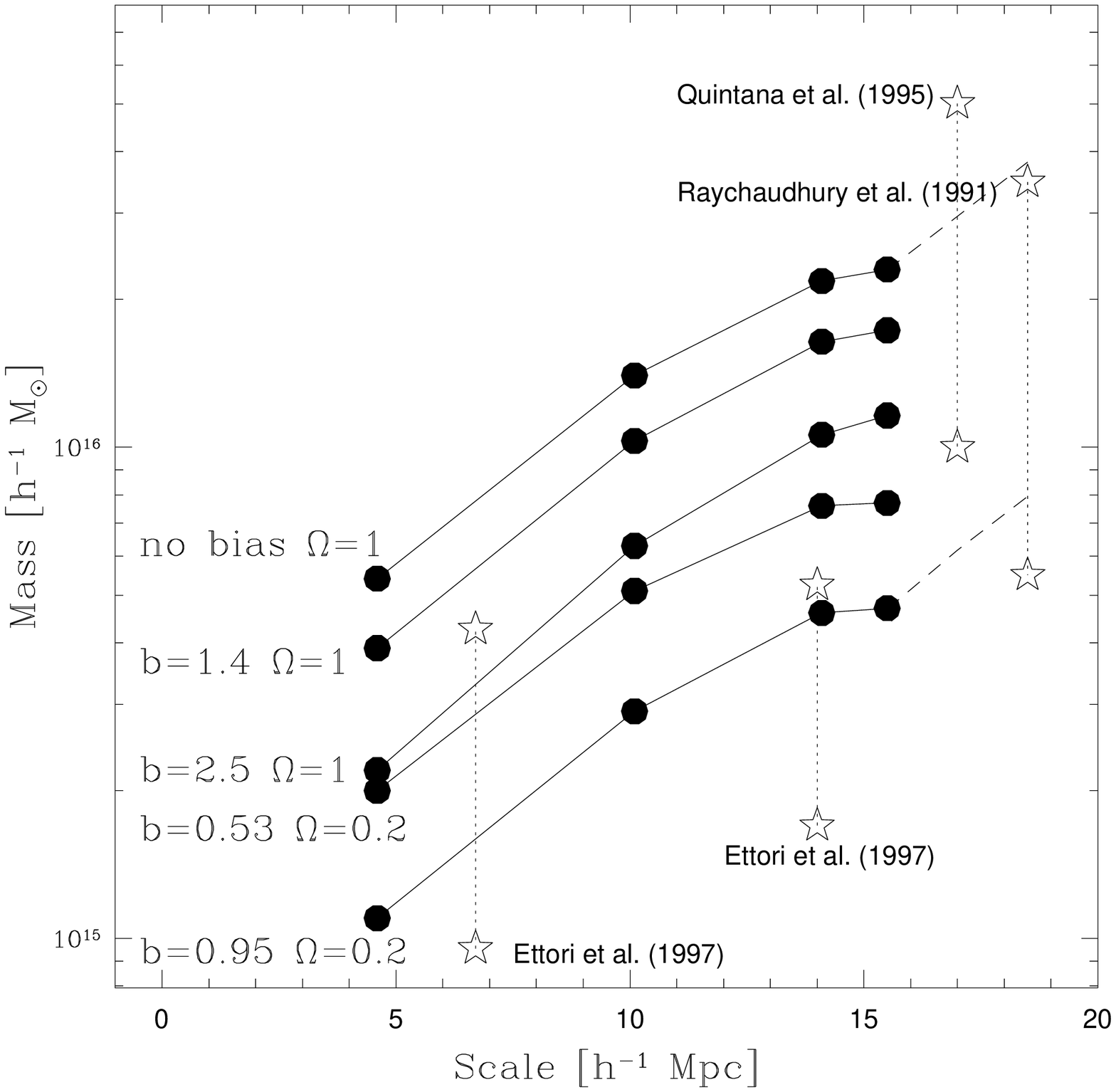} \hfil
\end{center}
Figure 5: Estimated masses at various scales for different parameter choices,
indicated near each curve.
Stars connected with dotted lines refer to mass ranges given in the literature.
Dashed lines give the extrapolation of our mass values to the scales of
Quintana et al. and Raychaudhury et al., assuming the same overdensity we found
at our largest scale.
\\
The overdensity and mass values for various sample are reported in the Table.
The case ``Total sample" includes all regions spectroscopically observed by us;
the case ``Extended sample" assumes that the estimated overdensity extends
also outside MEFOS fields, over the whole plates 443, 444 and 509; the case
``Extended sample $+$ 382 \& 383" is obtained adding also the overdensity values
found by Drinkwater et al. (1999) for plates 382 and 383. 
\end{figure}

%
\section{Spectral properties of galaxies}
{\bf In collaboration with A.Baldi}
\\
The large amount of spectra in our sample allows a uniform spectral
classification of galaxies in the Shapley Concentration, in order to
investigate the morphology--density relation in a wide range of densities, 
from the cluster to the supercluster environment, in this dynamically active 
region (Baldi, Bardelli \& Zucca, 2000).  
\\
The method we have chosen to perform spectral classification of our sample
of galaxies is the Principal Components Analysis, following the Galaz \&
deLapparent (1998) paper.
\\
We have considered only galaxies in the velocity range $10000 - 22500$ km/s, 
in order to limit the spectral analysis to the physical extension of the 
Shapley Concentration. Moreover, given the fact that the intercluster survey 
is limited to the the magnitude range 17 $\le$ $b_J$ $\le$ 18.8, when we 
compare the cluster galaxy properties with the intercluster ones, we use 
subsamples with these magnitude limits. Our final sample contains $\sim 800$
spectra.  
\\
Note that, although fiber spectra do not allow a very precise flux calibration,
in our case the use of a limited distance range maintains the spectral 
features at approximately the same wavelength and therefore at the same
instrumental response. 
\\
In order to have a comparison of supercluster spectra with those of ``true" 
field galaxies, we have considered a sample taken from
the ESO Slice Project (ESP) galaxy redshift survey (Vettolani et al. 1997).
This survey was obtained with the same telescope and instrumental set up
of our sample and therefore represents a well defined reference sample.
\\
From the PCA analysis, we found that the first three principal components (PCs)
(reported in Figure 6) are sufficient to reconstruct $\sim$99\% in flux
of the sample. Then, following Galaz \& deLapparent (1998), we applied a 
coordinate trasformation from the three-dimensional space defined by the PCs
to the bi-dimensional space $(\delta, \theta)$.
The $\delta$ parameter represents the importance of the blue part of the 
continuum with respect to the red one, while $\theta$ is an indicator of the 
emission line strength in a galaxy spectrum.
\\
The galaxies follow a well defined sequence in the $(\delta, \theta)$ plane
(Figure 6) and the Kennicutt's galaxies (of known morphological type) are 
located in a succession of increasing values of $\delta$, going from the 
ellipticals to the irregulars. For this reason, in the following we'll use 
$\delta$ to classify galaxies. 
\\
In Figure 7 the fractions of galaxies in three morphological bins (early,
intermediate and late type) are shown for each sample. 
As expected, cluster galaxies show a completely different morphological
distribution with respect to ESP and intercluster galaxies, being 
the fraction of early-type galaxies dominant in such high density environment.
\\
More interesting, this figure shows a broad agreement between the intercluster
and the ESP sample: this fact suggests that the galaxies located in the 
intercluster regions of the Shapley supercluster have a morphological 
mix not different from field galaxies. 
\\
Looking at the dependence of the morphological mix on the local density 
(Figure 7), a decreasing of the late-type galaxies fraction with increasing 
values of density is clearly visible, both in the A3528 and in the A3558 
complexes, reflecting the well-known morphology--density relation.   
Moreover, for every density value, this fraction remains significantly below 
the mean field value.
\\
From the A3528 complex, we have eliminated galaxies belonging to the poor 
cluster A3535 (which is not dynamically part of the structure): the 
morphological mix of these objects remains almost constant with the density 
and with values in agreement with those found for the field population.
\\
The spatial distribution of early-type and late-type galaxies in the two
cluster complexes is shown in Figure 8. 
\\
In Figure 9 we show the dependence of the mean [OII]
equivalent width and of the mean star formation rate (derived following
Kennicut 1992) on the local density.  

\begin{figure}
\begin{center}
\leavevmode
\epsfxsize=\hsize
\epsfbox{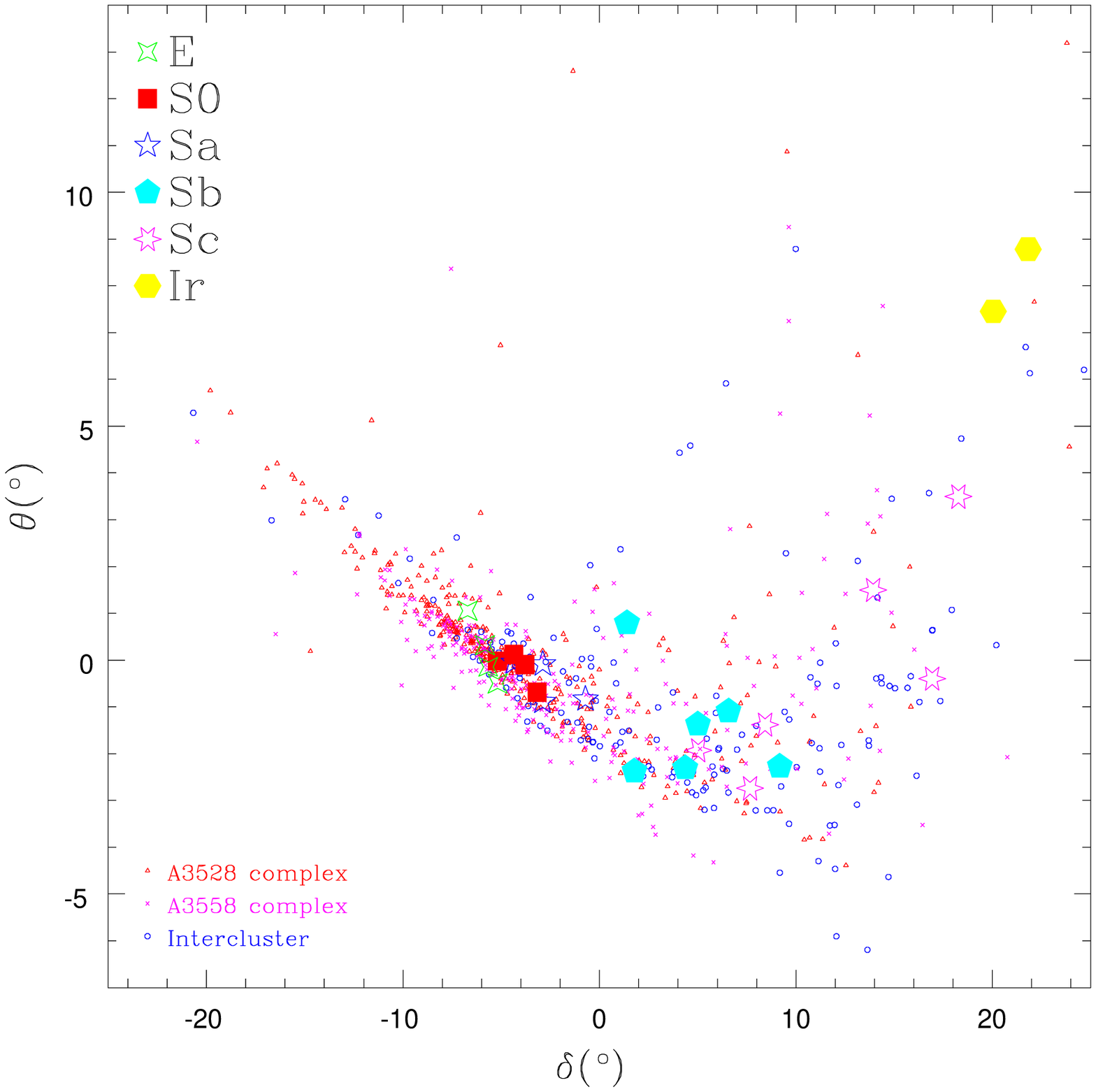} \hfil
\epsfxsize=0.32\hsize
\epsfbox{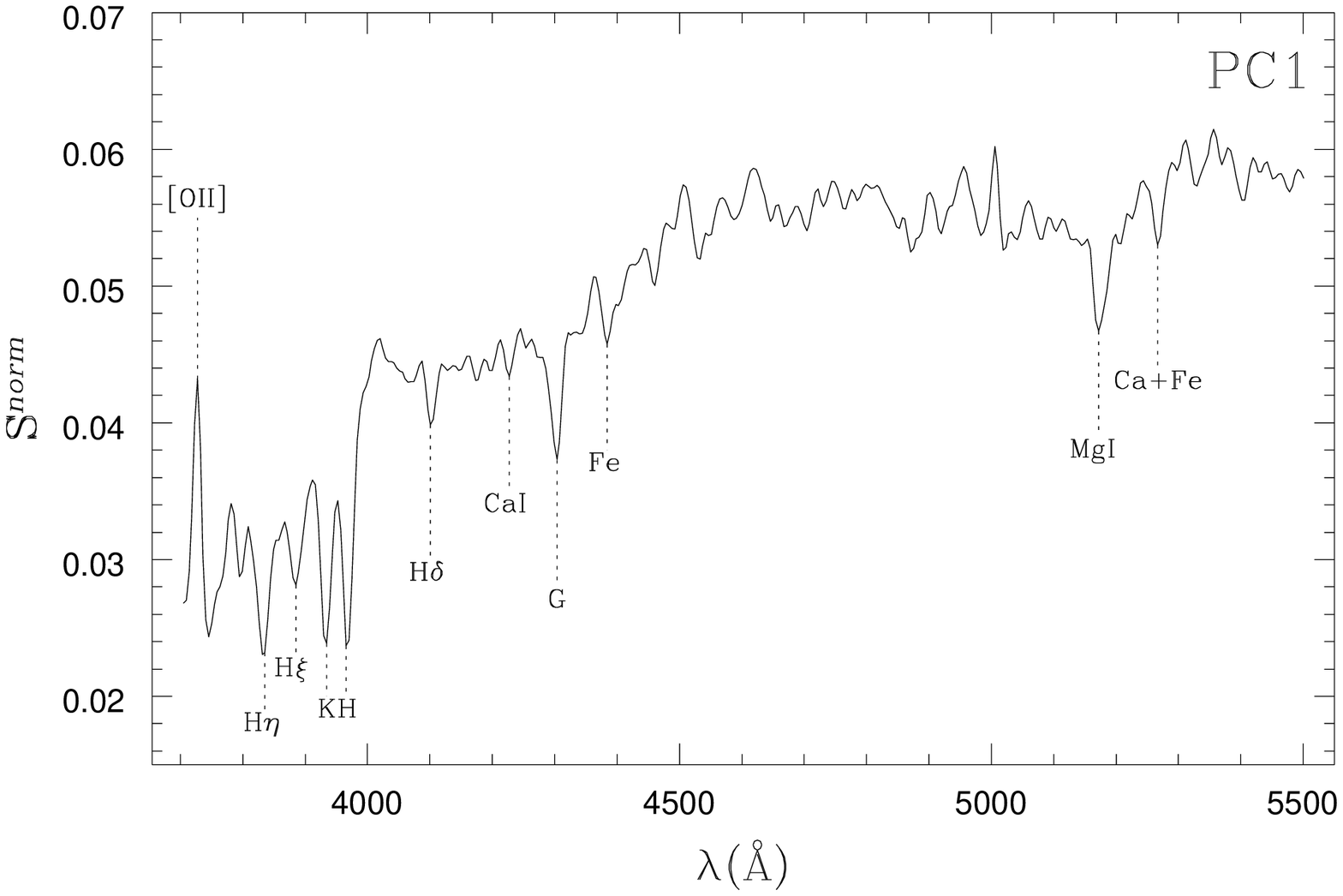} \hfil
\epsfxsize=0.32\hsize
\epsfbox{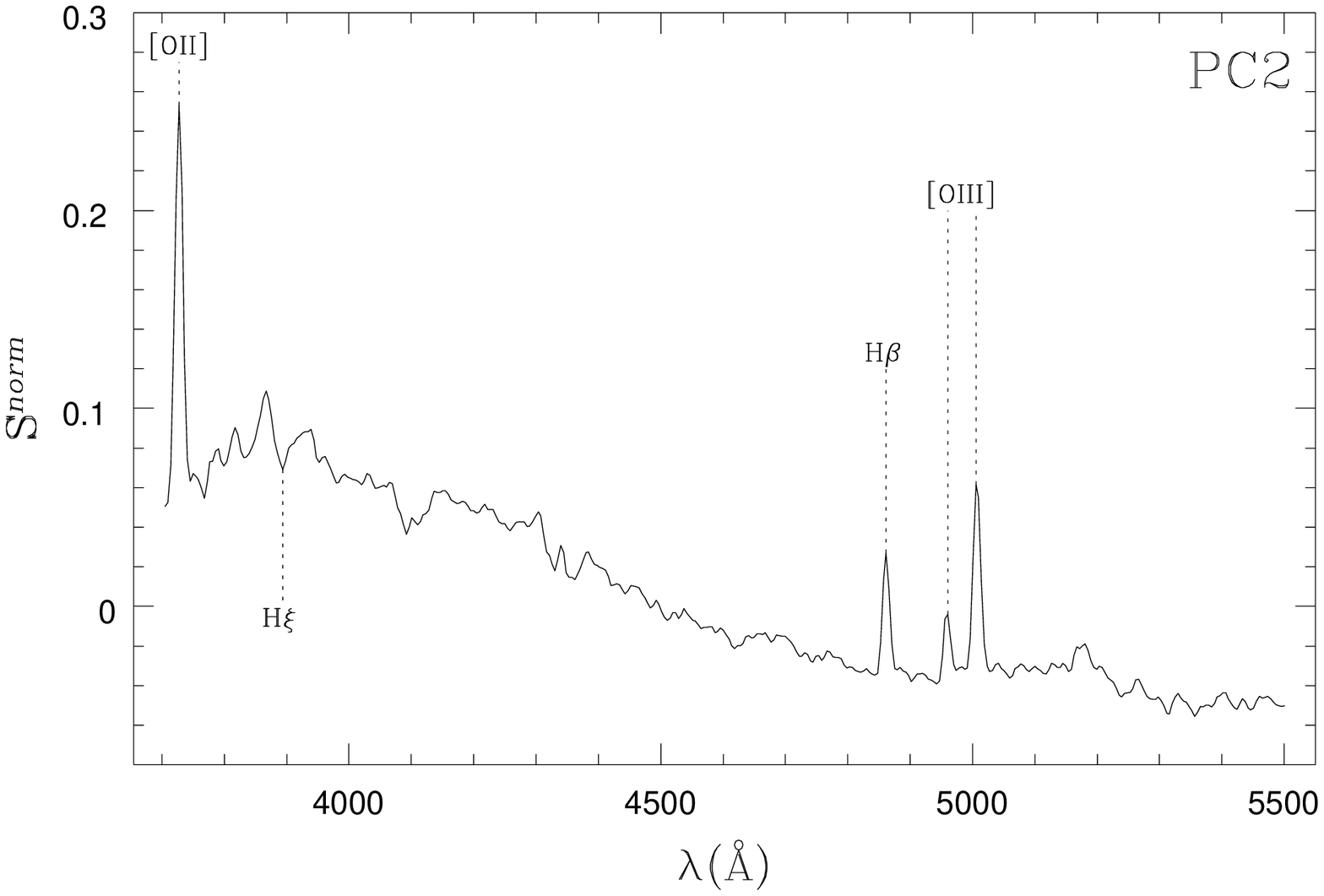} \hfil
\epsfxsize=0.32\hsize
\epsfbox{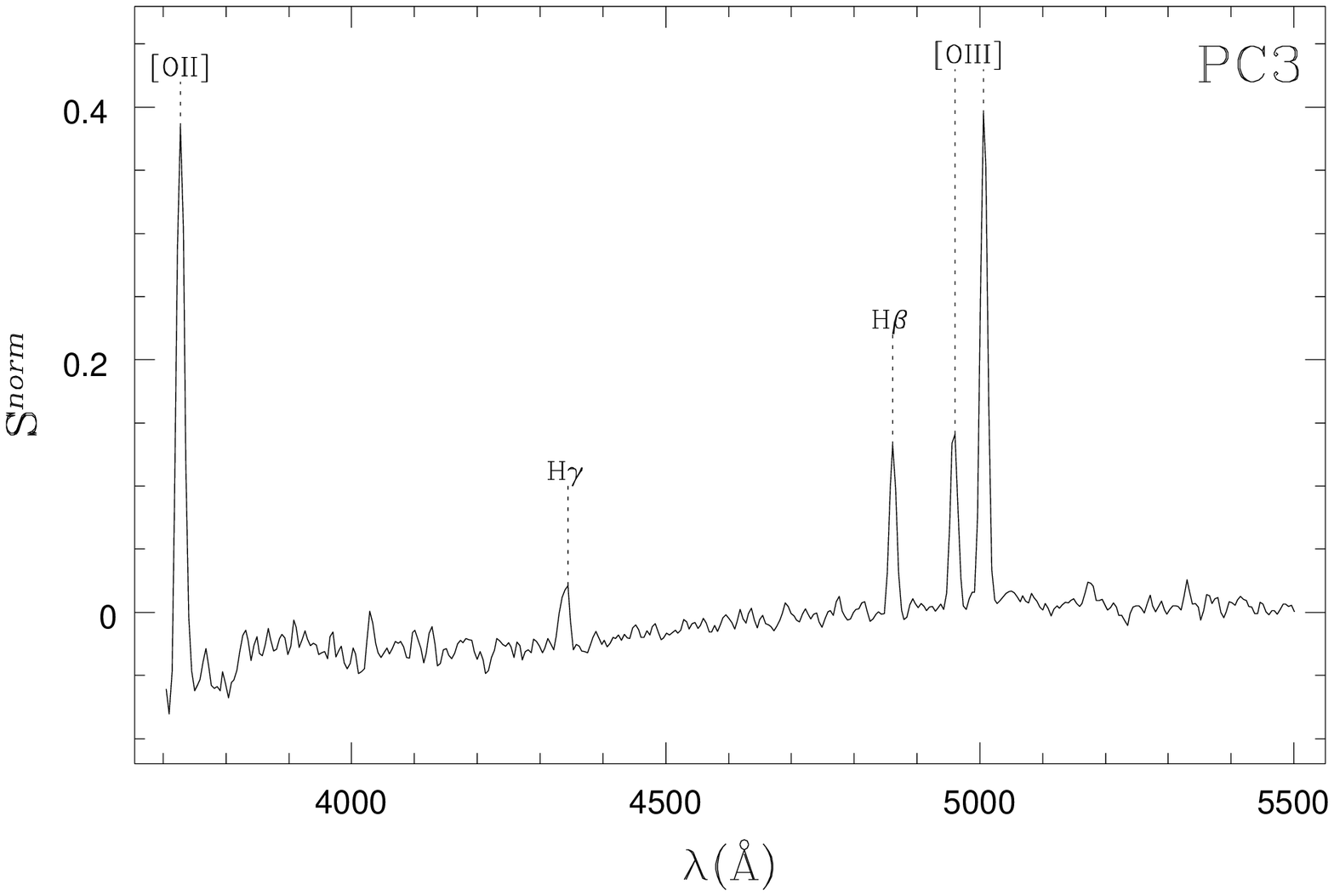} \hfil
\end{center}
Figure 6: The classification diagram $\delta$-$\theta$ for the spectral sample
in the Shapley Concentration plus the Kennicutt's sample (big symbols). 
The first three principal components (PCs) obtained from the PCA analysis
are shown in the lower panel.  
\end{figure}

\begin{figure}
\begin{center}
\epsfysize=10cm
\epsfbox{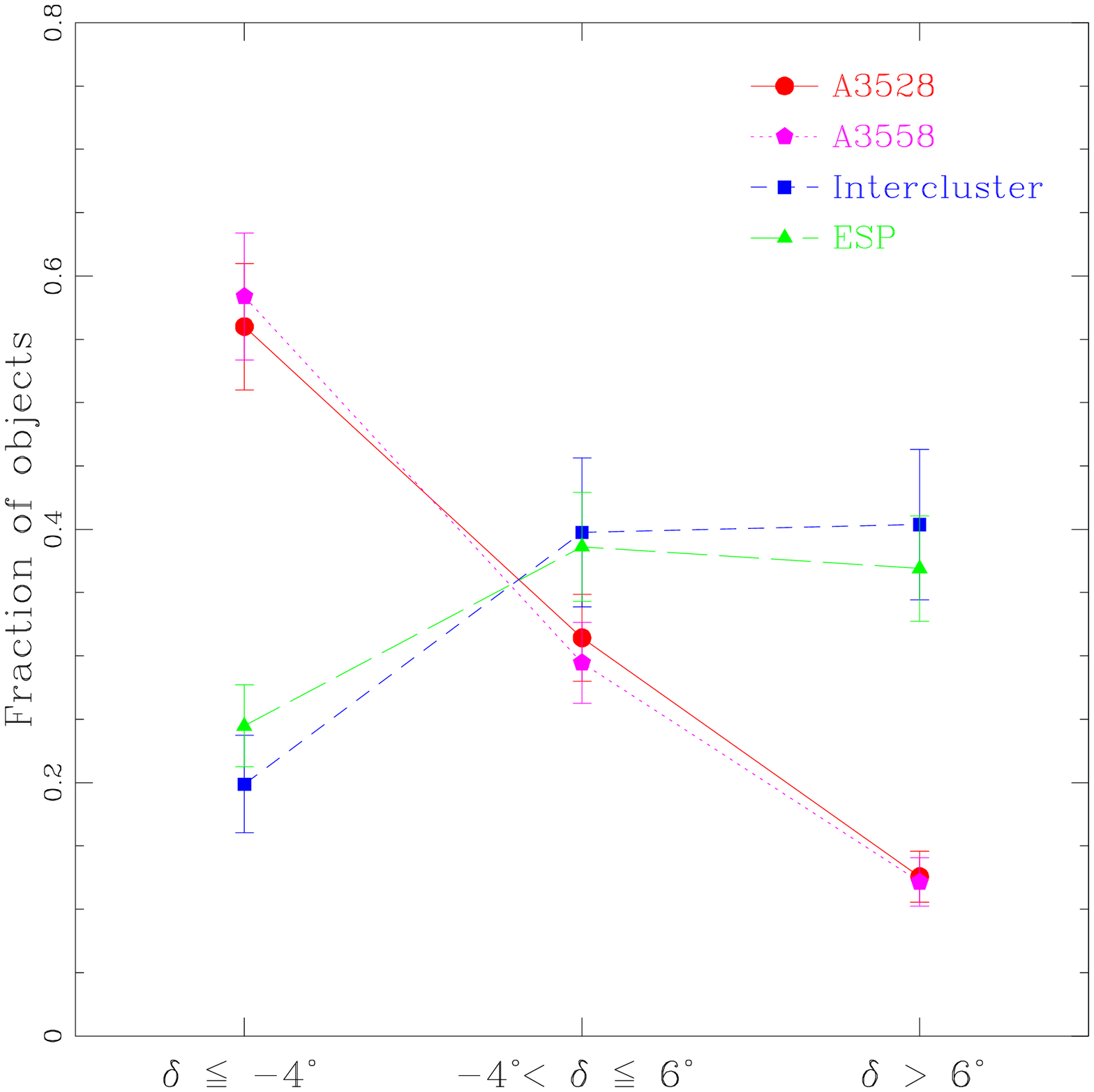}
\epsfysize=10cm
\epsfbox{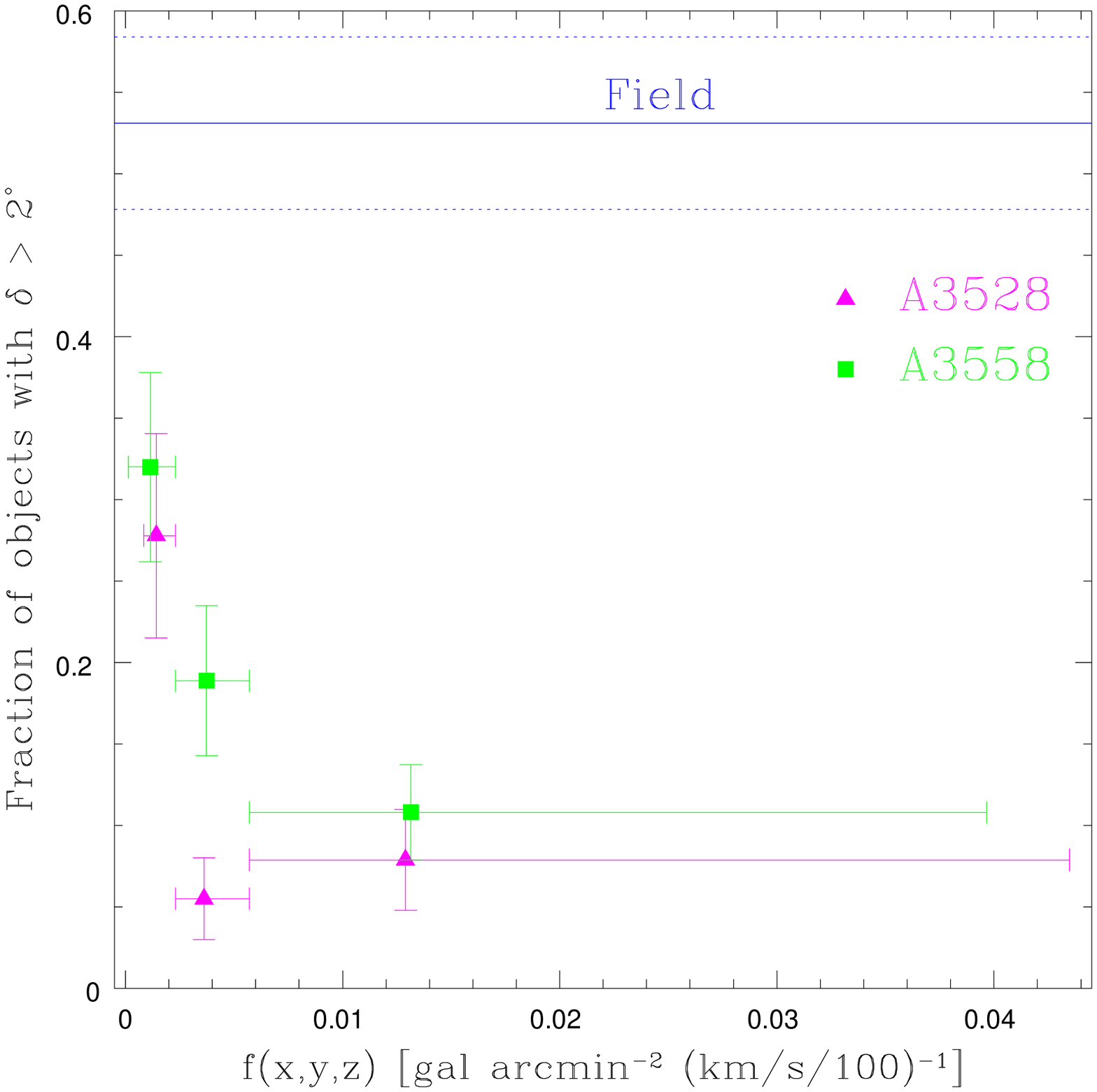}
\end{center}
Figure 7: Upper panel: Fractions of galaxies divided in three morphological 
classes, for each spectral sample: A3528 (circles), A3558 (pentagons), 
intercluster (squares) and ESP (triangles). Lower panel: 
Relation between the fraction of late-type galaxies and local 
density $f(x,y,z)$ in the A3528 (triangles) and A3558 (squares) complexes,
in the magnitude range $17.0 \le b_J \le 18.8$.  
Horizontal bars indicate the extension of the density intervals:
the points are located at the position of the density average value within
the bin. Vertical bars represent the 1-$\sigma$ uncertainties. 
The reference value for the field (from ESP galaxies) is reported as a solid
line: dotted lines represent its 1-$\sigma$ uncertainties. 
\end{figure}

\begin{figure}
\begin{center}
\epsfysize=14.5cm
\epsfbox{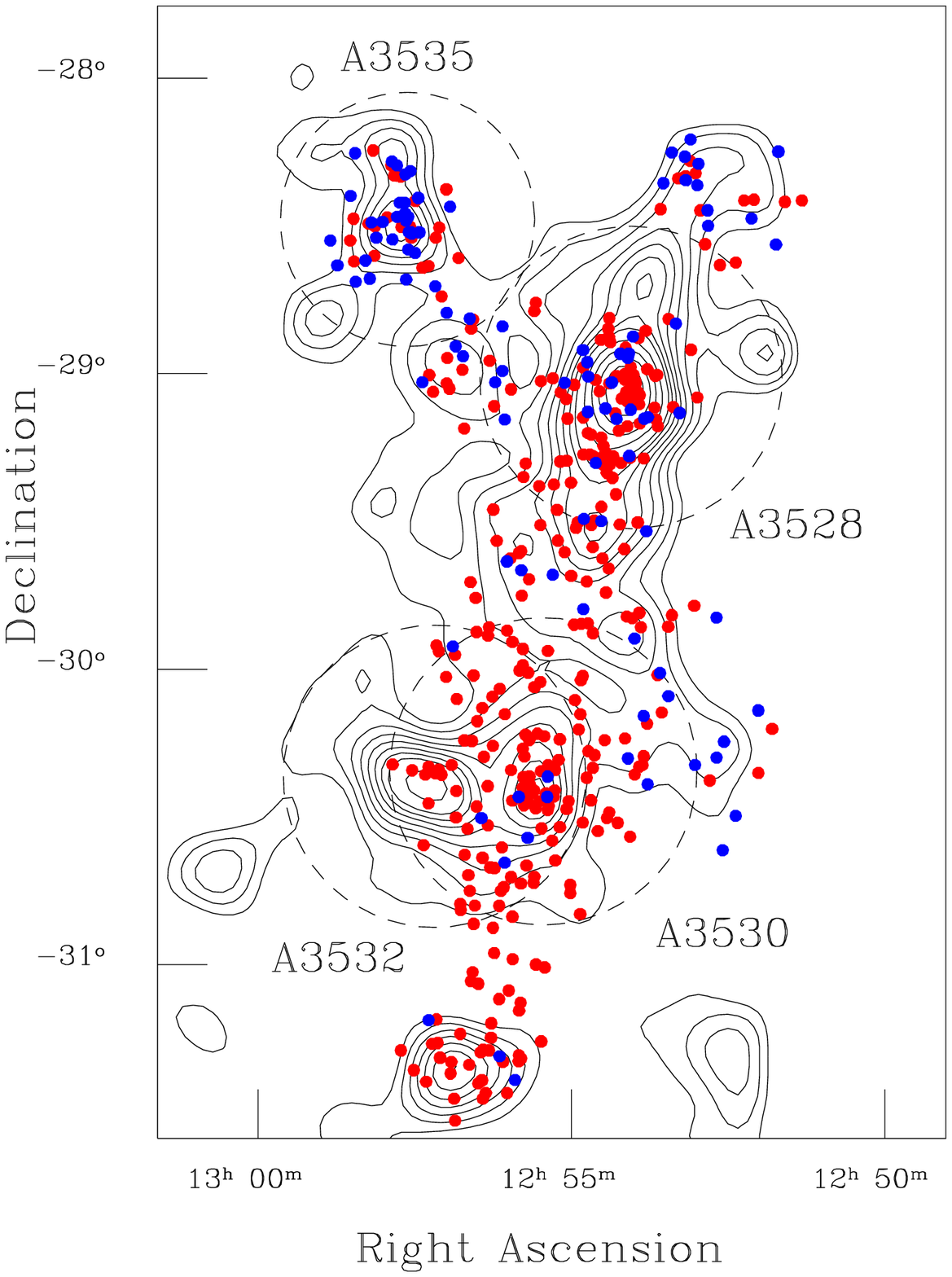}
\epsfxsize=0.9\hsize
\epsfbox{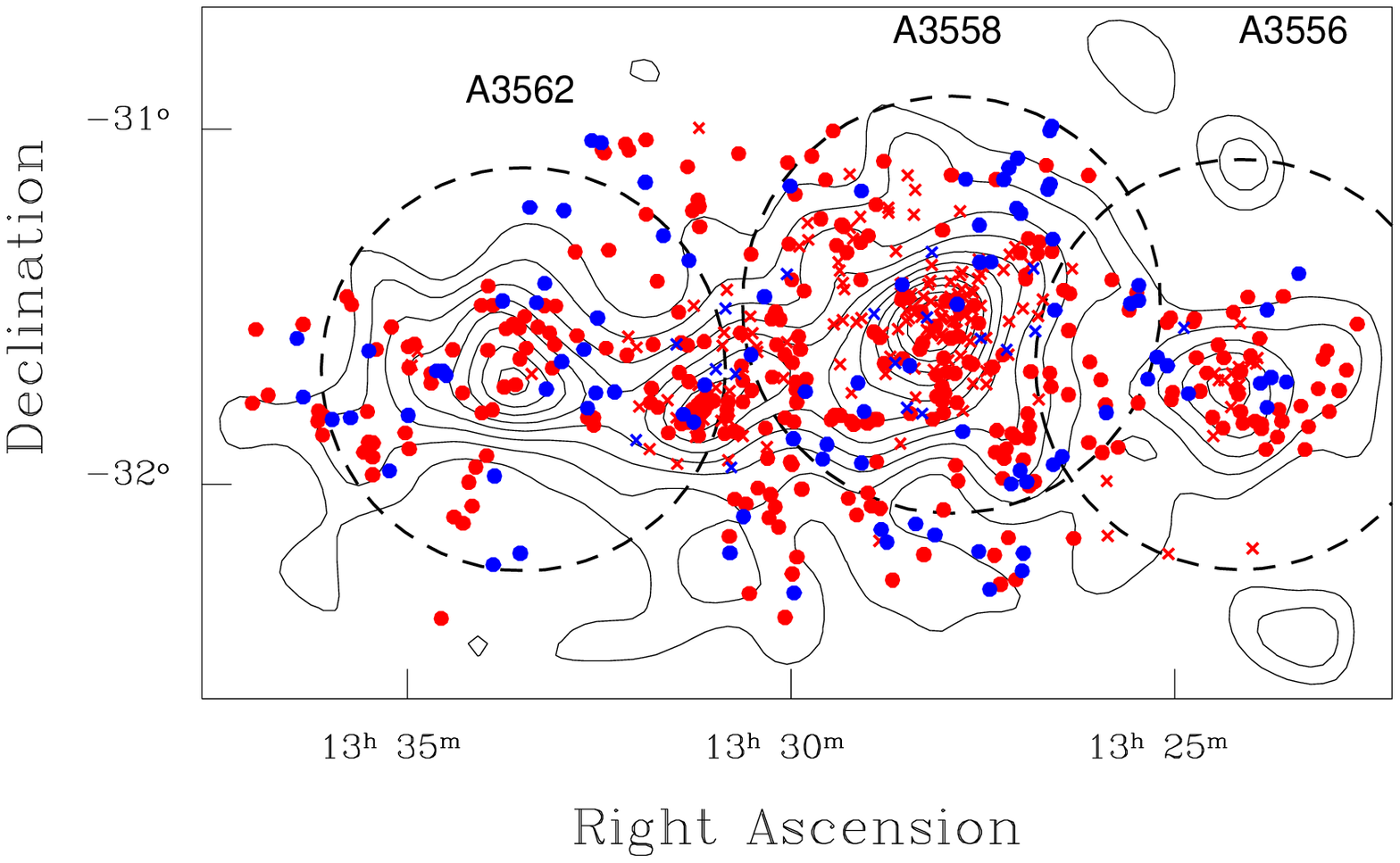}
\end{center}
Figure 8: Spectral classification of galaxies superimposed to the isodensity
contours in the cluster complexes. Red dots: early-type galaxies. Blue dots:
late-type galaxies. Upper panel: A3528 complex. Lower panel: A3558 complex
(in this panel, crosses indicate objects for which the classification has
been derived from literature $(U-B)$ colors). 
\end{figure}

\begin{figure}
\begin{center}
\epsfysize=10cm
\epsfbox{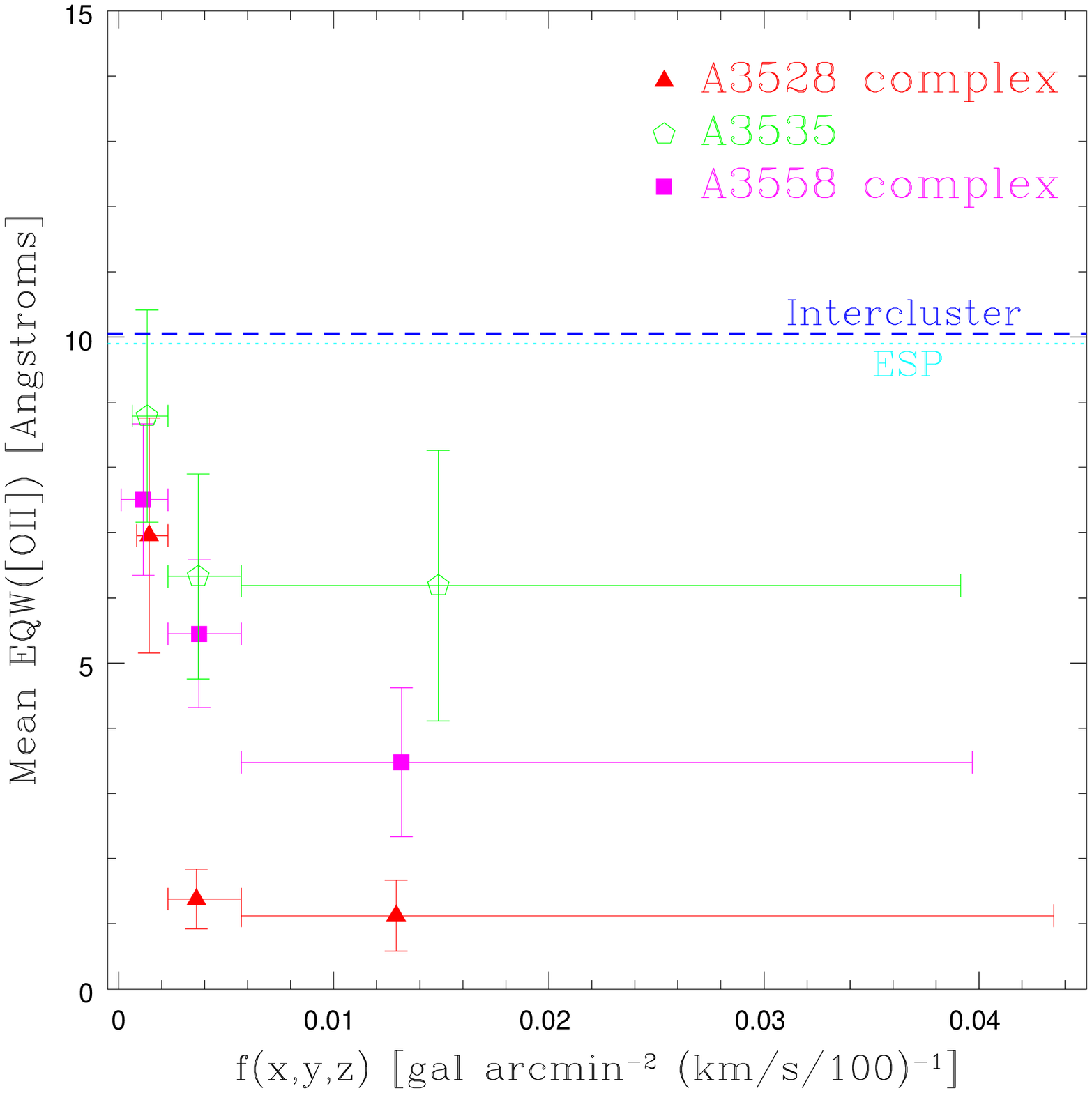}
\epsfysize=10cm
\epsfbox{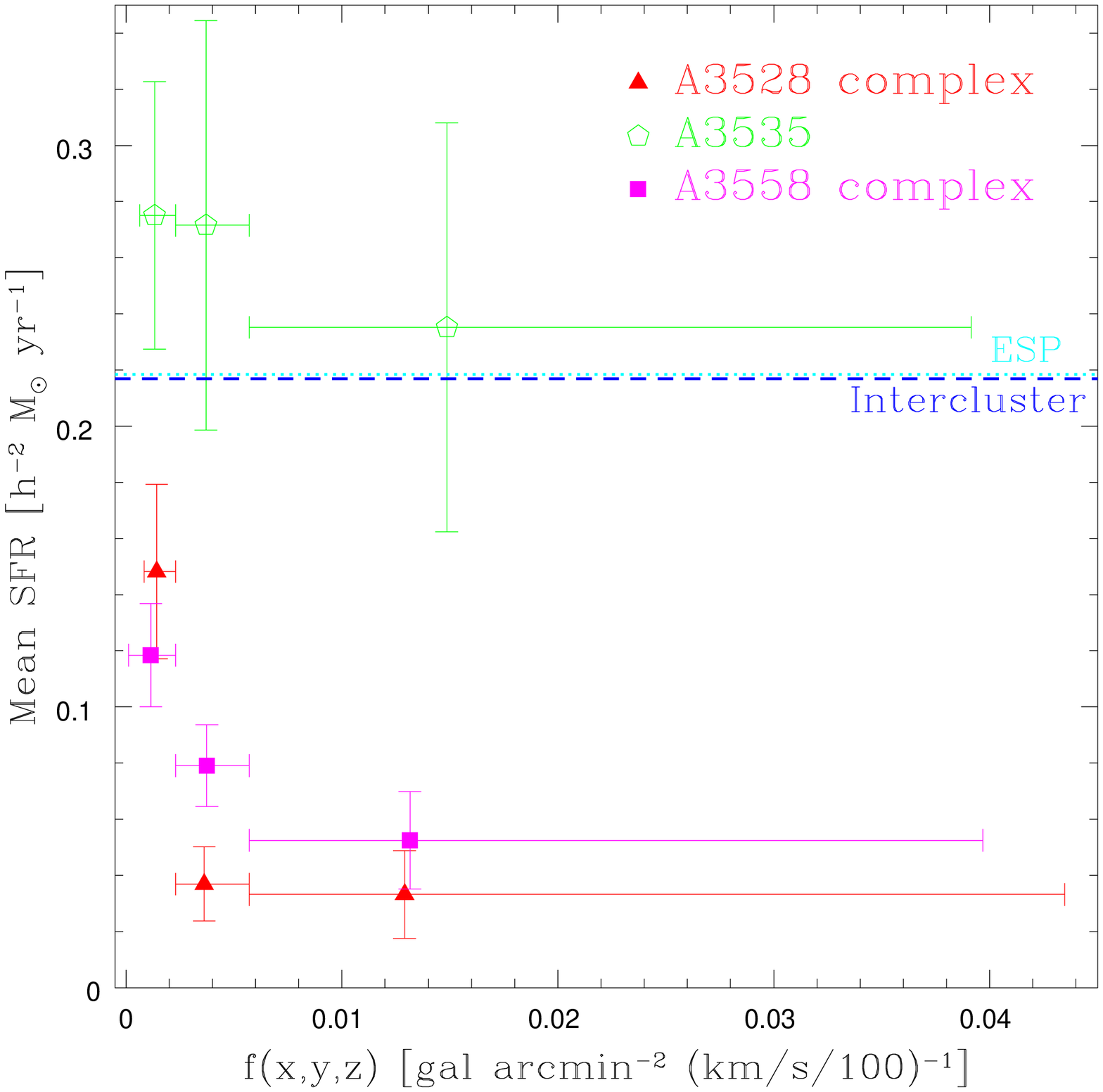}
\end{center}
Figure 9: Upper panel: Relation between the mean [OII] equivalent width and 
the local density $f(x,y,z)$ in the A3528 (triangles) and A3558 (squares) 
complexes, in the magnitude range $17.0 \le b_J \le 18.8$. The meaning of
the intervals is the same as Figure 2. The A3535 cluster has been considered
separately, because it is not dynamically part of the A3528 complex.
The horizontal blue line represents the reference value for the field.
Lower panel: Relation between the mean star formation rate (derived following
Kennicut 1992) and the local density $f(x,y,z)$, for the same samples.  
\end{figure}

\end{document}